%% file: ms.tex
\documentclass[preprint2]{aastex}    

\usepackage{longtable}
\usepackage{lscape}

\newcommand\LSIV{LS\,IV$-08^\circ 3$}
\newcommand\JKs{\ensuremath{J\!-\!K_S}}
\newcommand\HKs{\ensuremath{H\!-\!K_S}}
\newcommand\JH{\ensuremath{J\!-\!H}}
\newcommand\CIII{\ion{C}{3}}
\newcommand\NIII{\ion{N}{3}}
\newcommand\HI{\ion{H}{1}}
\newcommand\HeI{\ion{He}{1}}
\newcommand\HeII{\ion{He}{2}}
\newcommand\Ha{H$\alpha$}
\newcommand\Hb{H$\beta$}
\newcommand\NaI{\ion{Na}{1}}
\newcommand\OI{\ion{O}{1}}

\slugcomment{to be published in Astronomical Journal}

\shorttitle{A New Bright Emission Line Binary}
\shortauthors{Stark et.\ al.}

\begin{document}

\title{A New, Bright, Short-Period, Emission Line Binary in
Ophiuchus\footnote{Based in part on observations obtained with the
Hobby-Eberly Telescope, which is a joint project of the University of
Texas at Austin, the Pennsylvania State University, Stanford
University, Ludwig-Maximillians-Universit\"at M\"unchen, and
Georg-August-Universit\"at G\"ottingen.}}

\author{M.\ A.\ Stark\footnote{Visiting Astronomer, Kitt Peak National
Observatory, National Optical Astronomy Observatory, which is operated
by the Association of Universities for Research in Astronomy,
Inc. (AURA) under cooperative agreement with the National Science
Foundation.}}
\affil{Department of Physics and Astronomy, University of Wyoming,
Dept.\ 3905, 1000 E.\ University Ave.,\ Laramie, WY 82071}
\email{stark@uwyo.edu}

\author{Richard A.\ Wade}
\affil{Department of Astronomy and Astrophysics, The Pennsylvania
State University, 525 Davey Lab, University Park, PA 16802}
\email{wade@astro.psu.edu}

\author{John R.\ Thorstensen, Christopher S.\ Peters}
\affil{Department of Physics and Astronomy,
6127 Wilder Laboratory, Dartmouth College,
Hanover, NH 03755-3528}

\author{Horace A.\ Smith, Robert D.\ Miller}
\affil{Department of Physics and Astronomy,
Bio-Physical Sciences Bldg, Michigan State University,
East Lansing, MI 48824-1116}

\author{E.\ M.\ Green} \affil{Steward Observatory, University of
Arizona, Tucson, AZ 85721-0065}

\begin{abstract}
The 11th magnitude star {\LSIV} has been classified previously as an
OB star in the Luminous Stars survey, or alternatively as a hot
subdwarf.  It is actually a binary star.  We present spectroscopy,
spectroscopic orbital elements, and time series photometry, from
observations made at the Kitt Peak National Observatory 2.1m, Steward
Observatory 2.3m, MDM Observatory 1.3m and 2.4m, Hobby-Eberly 9.2m,
and Michigan State University 0.6m telescopes.  The star exhibits
emission of varying strength in the cores of H and {\HeI} absorption
lines.  Emission is also present at 4686\,\AA\ ({\HeII}) and near
4640/4650\,\AA\ ({\NIII}/{\CIII}).  Time-series spectroscopy collected
from 2005 July to 2007 June shows coherent, periodic radial velocity
variations of the {\Ha} line, which we interpret as orbital motion
with a period of 0.1952894(10)\,days.  High-resolution spectra show
that there are two emission components, one broad and one narrow,
moving in antiphase, as might arise from an accretion disk and the
irradiated face of the mass donor star. Less coherent, low-amplitude
photometric variability is also present on a timescale similar to the
orbital period.  Diffuse interstellar bands indicate considerable
reddening, which however is consistent with a distance of
$\sim$100--200 pc.  The star is the likely counterpart of a weak {\em
ROSAT} X-ray source, whose properties are consistent with accretion in
a cataclysmic variable (CV) binary system.  We classify {\LSIV} as a
new member of the UX UMa subclass of CV stars.
\end{abstract}



\section{Introduction} \label{sec:Intro}

The star {\LSIV} has $V = 11.475$, $(\bv) = +0.190$ {\citep{RN00}},
and was first classified ``OB'' in the {\em Luminous Stars in the
Northern Milky Way} catalog {\citep{LSIV}}.  Later it was reclassified
``sdB'' by {\citet{KB92}} based on Reticon spectrograms and
Str\"{o}mgren photometry.  Additional photometry of this object is
available in the literature {\citep{RCL,Reed,RN00}}, but no additional
spectroscopic observations have been reported. Astrometric and
photometric data for this star, collected from the literature,
are reported in Table~\ref{table:data}.

We obtained spectra of {\LSIV} as part of our program on composite
spectrum hot subdwarf stars {\citep{SW,SW04,Thesis,SW06}}, owing to its
unusually red $V-K_s$ and $J-K_s$ colors as observed in the {\em Two
Micron All-Sky Survey} {\citep[2MASS;][]{Skrutskie06}}.  We found that
{\LSIV} exhibits broad absorption lines with core emission, which is
variable in both strength and profile shape.  We obtained time-series
photometry and additional spectroscopy in order to correctly classify
this star.


\section{Observations} \label{sec:Observations and Findings}

\subsection{Low-Dispersion Spectra}\label{sec:Low-disp}

\subsubsection{Kitt Peak National Observatory 2.1m Telescope}\label{sec:GoldCam}

\begin{figure}
\epsscale{0.45}
{\plotone{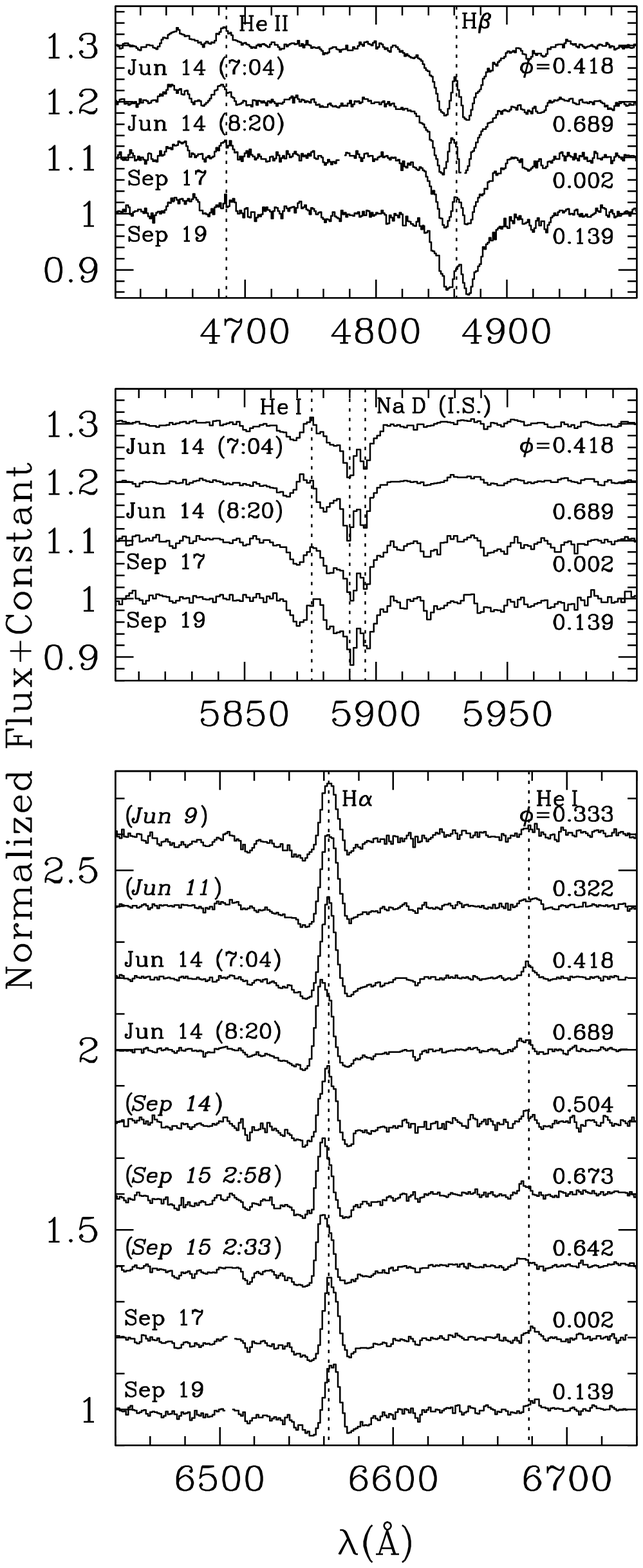}}
\caption{KPNO 2.1m Goldcam spectra of {\LSIV} showing the variable
nature of the emission components at {\Hb} ({\em top}),
{\HeI}~5875\,\AA\ partly blended with interstellar {\NaI}~D lines
({\em middle}), and {\Ha}/{\HeI}~6678\,\AA\ ({\em bottom}). Dates and
times are UT 2003; times correspond to mid-exposure and are
heliocentric.  Phases corresponding to the ephemeris in \S\ref{sec:RV}
are also indicated.  Dates that are italicized in parentheses in the
bottom panel do not have corresponding data for {\Hb} and
{\HeI}~5875\,\AA.  Dotted lines show the laboratory wavelengths of the
labeled transitions; the spectra themselves are plotted using observed
(topocentric) wavelengths. Several gaps in the spectra result from
cosmic ray removal.
\label{fig:GoldCamspec}}
\end{figure}

{\LSIV} was observed from the Kitt Peak National Observatory (KPNO)
2.1m telescope using the GoldCam spectrograph during 2003 June and
September.  A journal of observations is given in
Table~\ref{table:GoldCamObs}.  All exposures were between 5 and 20
minutes long. Two different spectrograph configurations were used,
which covered approximately 4600--6750\,\AA\ and 6400--8900\,\AA\ at
$\sim$1.3\,\AA~pixel$^{-1}$.  The features observed in the GoldCam
spectra include broad absorption lines with emission cores in {\Ha},
{\Hb}, and {\HeI} 5875\,\AA; emission from {\HeI} 6678\,\AA, {\HeII}
4685\,\AA, and {\NIII}/{\CIII} $\sim$4640/4650\,\AA.  The emission
components were observed to vary in both strength and profile in as
little as $\sim$30 minutes (Fig.~\ref{fig:GoldCamspec}; see also
\S\ref{sec:HRS}).

All GoldCam spectra were processed using standard IRAF\footnote{IRAF
is distributed by the National Optical Astronomy Observatories, which
are operated by the Association of Universities for Research in
Astronomy, Inc., under cooperative agreement with the National Science
Foundation.}  routines.  Unfavorable observing conditions prevented
reliable flux calibrations on most nights, so we discuss only
the normalized spectra.  The continua were fitted by hand using
splines.  The equivalent widths (EWs) of features of interest were
measured in each spectrum; the wavelength interval encompassed by each
feature was determined by eye, noting where the line deviated from the
continuum.  For lines with broad absorption wings and core emission,
two EW measurements were made.  The first is the total EW of the whole
line (absorption and emission included).  The second EW refers to just
the emission component, measured above the point where the emission
emerges from the absorption wings, giving a lower limit for the
``true'' emission EW.  Reported in Table~\ref{table:EWLSIV} are the
average EWs and their formal errors (not including any systematic
errors or accounting for variability). Also reported are the number of
measurements, their standard deviation, minimum and maximum values,
and the difference between the maximum and minimum. (Note that EW$>0$
indicates absorption, while EW$<0$ indicates emission.)

Significant reddening of this object is evidenced by strong
interstellar {\NaI}~D lines and several diffuse interstellar bands
(DIBs).  {\citet*{Schlegel}} report total galactic reddening of
$E(\bv) \approx +0.48$ along the line of sight to {\LSIV}. The
reddening and distance are discussed in \S\ref{sec:reddening}.

\subsubsection{Steward Observatory 2.3m Telescope} \label{sec:Steward}

\begin{figure}
\epsscale{0.40}
\rotatebox{-90}
{\plotone{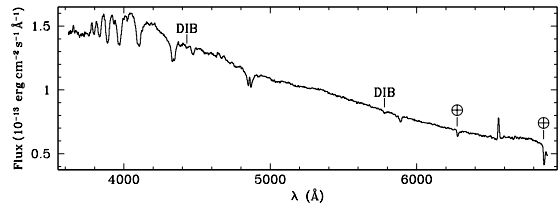}}
\caption{Flux-calibrated spectrum of {\LSIV}, obtained UT 2002 April 3
at the Steward Observatory 2.3m telescope, including higher-order
Balmer lines.  Absorption bandheads at $\lambda$6277 and $\lambda$6867
are telluric; absorption features at $\lambda$4430 (near the stronger
\ion{He}{1} $\lambda$4471 feature) and $\lambda$5780 are diffuse
interstellar bands.
\label{fig:betsyflux2002}}
\end{figure}

A flux-calibrated spectrum of {\LSIV}, covering 3620--6895\,\AA\ at
$\sim$3\,\AA~pixel$^{-1}$ dispersion ($\sim$9\,\AA\ resolution), was
obtained on UT 2002 April 3 at the Bok 2.3m telescope on Kitt Peak,
Arizona, using the B\&C spectrograph with a thinned, back-illuminated
1200 $\times$ 800 Loral CCD.  This ``pre-discovery'' spectrum is shown
in Figure~\ref{fig:betsyflux2002}.  It extends to
shorter wavelengths then the GoldCam observations, including the Balmer
lines up to the series limit.  A second spectrum was obtained with the
same instrument setup on UT 2007 May 8 and was flux-calibrated in a
homogeneous way with the first; the overall spectral energy
distribution is very similar, with the flux level differing by about
3\%.  The H emission lines were weaker in 2007, with the emission
reversal in H$\gamma$ virtually disappearing and the height of {\Ha}
emission above the continuum diminishing by a factor of 2.
\citet{KB92} classified {\LSIV} as ``sdB'' based on {\em uvby}
photometry and 2\,\AA\ resolution Reticon spectroscopy covering
3900--4600\,\AA, so {\Hb} was not observed, and weak emission in the
absorption core of H$\gamma$ might easily have been overlooked or even
absent.  The spectral energy distribution (SED) is quite blue, even
despite the reddening evidenced by the DIBs. The SED is further
discussed in \S\ref{sec:Discussion}.


\subsubsection{MDM Observatory 1.3m and 2.4m Telescopes} \label{sec:MDM}

Time-series spectra of {\LSIV} were obtained at the MDM
Observatory\footnote{MDM Observatory is operated by a consortium
consisting of Dartmouth College, Columbia University, Ohio State
University, the University of Michigan, and Ohio University.}  on Kitt
Peak, Arizona.  Table~\ref{table:mdmjournal} gives a journal of
observations.  Most of the MDM spectra were obtained using the Mark
III spectrograph and a SITe 1024$^2$ CCD detector mounted on the
McGraw-Hill 1.3m telescope.  A 600 line mm$^{-1}$ grism gave
$\sim$2.3\,\AA\,pixel$^{-1}$ dispersion over the range
4660--6900\,\AA, with a full width half maximum (FWHM) resolution of
4.1\,\AA.  All stellar exposures were bracketed with spectra of Hg,
Ne, and Xe comparison lamps, in order to track the spectrograph
flexure.  (In 2004 June, an attempt was made to use the $\lambda 5577$
night-sky emission feature to set the wavelength zero point;
unfortunately, this did not work well, and those spectra are excluded
from the present analysis.)

A few MDM spectra are from the Hiltner 2.4m telescope and modular
spectrograph, equipped with a SITe 2048$^2$ CCD; this gave 2\,\AA\
pixel$^{-1}$ dispersion from 4300 to 7500\,\AA, with severe vignetting
toward the ends of this range.  The spectral resolution of this setup
was slightly better than that of the Mark III.  Some of these
exposures were taken in bright twilight and bracketed with comparison
exposures; for those taken away from twilight, the night-sky features
were used to find the zero point shift.  Extensive internal checks
show that this procedure worked well with this setup.  Aliasing
related to daily cycle count uncertainties was suppressed by pushing
to large hour angles; the target's brightness made it possible to
obtain good signal-to-noise spectra even at large airmass.

Flux standard stars were observed to calibrate the instrument
response, when the twilight sky was reasonably clear.  Experience
suggests that the zero point of the MDM spectra should be accurate to
about 20\%.  Exposures of bright O and B stars were also obtained and
used to map and correct for the telluric absorption features.

An average flux-calibrated spectrum was constructed from the 1.3m data
taken in 2005 June and July.  Using the IRAF {\it sbands} task and the
passband tabulated by \citet{bessell}, we find a synthetic $V$
magnitude of 11.47.  The MDM 1.3m average spectrum is very similar to
the Steward 2.3m spectrum from 2002 April in the wavelength region
common to both, except the overall flux level is about 10\% higher,
and the emission lines are more evident (the {\Hb} emission reversal
protrudes slightly above the continuum; the \ion{He}{1} $\lambda$6678
line is in emission at about 15\% of the intensity of {\Ha}).

\subsubsection{Radial Velocity Variations} \label{sec:RV}

All of the MDM spectra show {\Ha} emission.  To measure the radial
velocity (RV), we convolved the line profile with the derivative of a
Gaussian and took the line center to be the zero of this convolution
\citep{sy80}.  The width of the convolving function was optimized for
a line with a FWHM of 11.5\,\AA, which is approximately the observed
line width.  We searched the velocity time series for periodicities
using a ``residual-gram'' technique \citep{tpst}, which is especially
effective when the modulation is accurately sinusoidal.  Because the
data span hundreds of days, we were careful to use a sufficiently fine
grid of trial frequencies.  The period search yielded a single,
uniquely-defined frequency near 5.12 cycle d$^{-1}$; a least-squares
fit to the velocities of the form
$$v(t) = \gamma + K \sin\left[2\pi (t - T_0) \over P\right]$$ then
yielded a preliminary period of 0.195290 d, with a velocity
semiamplitude of 80~km~s$^{-1}$.

\begin{figure}
\epsscale{0.55}
\plotone{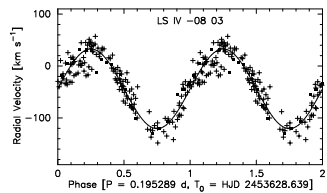}
\caption{{\Ha} radial velocities used in the analysis, folded on the
best-fitting period, with the sinusoidal fit superposed.  {\em
Crosses:} MDM velocities.  {\em Filled squares:} HET-HRS velocities.
Two cycles are shown for continuity.
\label{fig:foldedRV}}
\end{figure}

Twenty-three additional RVs were used to extend the span of RV
observations and to improve the precision of the ephemeris.  These
were measured using the same 11.5\,\AA\ FWHM Gaussian-derivative
technique, applied to high-resolution spectra from the Hobby-Eberly
Telescope (HET) obtained in 2004 and 2007 (see \S\ref{sec:HRS} below).
The final heliocentric {\Ha} RV ephemeris, based on 163 points, is
given by:
\begin{eqnarray*}
T_0 &=& \hbox{HJD } 2453628.6389 \pm 0.0016, \\
  P &=& 0.1952894 \pm 0.0000010\ \hbox{d}, \\
  K &=& 77 \pm 4\ {\rm km\ s^{-1}}, \ \hbox{and} \\
  \gamma &=& -44 \pm 3\ {\rm km\ s^{-1}}. \\
\end{eqnarray*}
The rms residual between the data and the fitted curve is 17
km~s$^{-1}$.  Figure \ref{fig:foldedRV} shows the RVs folded on the
period, with the best-fitting sinusoid superposed.  The periodicity in
the {\Ha} RVs is coherent over three observing seasons, and the
velocities admit no possibility of a cycle count error --- the data
are consistent with a binary orbit.

\subsection{High-Resolution Spectra} \label{sec:HRS}

Follow-up observations were taken at the HET using the High Resolution
Spectrograph \citep[HRS;][]{Tull98}.  Owing to the declination of
{\LSIV}, it can only be observed by the HET for about one hour per
night as it is transiting, from mid-April through mid-July.  Queue
observations of {\LSIV} were made with the HRS using $2''$ fibers (one
object fiber and two sky fibers) and 2$\times$3 on-chip binning,
resulting in $R \equiv \lambda/\Delta\lambda \approx30,000$, covering
the wavelength region from {\Hb} to {\Ha}.  (Unfortunately, the
$\lambda$5780 DIB falls into the gap between the ``red'' and ``blue''
CCDs with this setup.)  Exposure durations were 10 minutes, with the
exception of the first observation of 20 min duration and a few
shortened exposures obtained as {\LSIV} approached the end of the HET
track.  Three or four separate exposures were obtained in quick
succession on each of four nights in 2004 and three nights in 2007
(see Table~\ref{table:HRS} for a journal of observations).  The typical
signal-to-noise ratio was 100 at 5822\,\AA\ (range 75--120). Th-Ar
lamps and flat field exposures were taken either before or after the
sequence of exposures each night.

\begin{figure}
\epsscale{0.45}
\rotatebox{-90}{\plotone{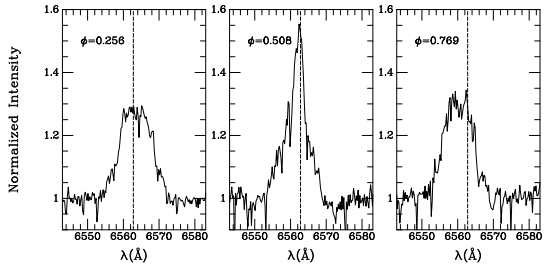}}
\caption{HET-HRS spectra showing only the emission component of the
{\Ha} line, for selected orbital phases: 
{\em left:} $\phi$=0.256 (2004 May 24); 
{\em middle:} $\phi$=0.508 (2007 April 19); 
{\em right:} $\phi$=0.769 (2004 May 29).  
Wavelengths are heliocentric.
The spectra have been ``stretched'' to make the emission EW uniformly
3.0\,\AA\ (see \S\ref{sec:HRS}).  
The variation of the emission profile is evident.  
The dashed vertical line in each panel indicates the laboratory
wavelength of {\Ha}.
The narrow absorption lines, which are stationary in the topocentric
frame, are telluric.
\label{fig:HRS-keyphases}}
\end{figure}

\begin{figure}
\epsscale{0.55}
\plotone{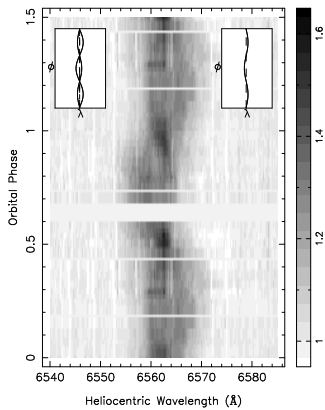}
\caption{A trailed-spectrogram representation of the variations in the
{\Ha} emission profile from the HET-HRS.
Wavelengths are heliocentric.  Orbital phase increases upwards, with
1.5 cycles being shown for continuity (data are repeated).
Intensities are in units of the continuum level and are indicated by
the calibration strip along the {\em right} side of the main panel.
The first inset ({\em upper left}) is intended as a guide to
identifying the two distinct emission components moving in antiphase.
The second inset ({\em upper right}) shows the single RV curve, with
reduced amplitude, resulting from the analysis of the low-resolution
MDM data.  See text for details and cautions.
\label{fig:trailedspec}}  
\end{figure}

These spectra were processed with standard IRAF routines.  At the
dispersion of the HRS, the absorption components of the Balmer lines
in the {\LSIV} spectrum are roughly as broad as a single spectral
order.  Thus, a pseudo-continuum was fitted to the broad H absorption
profiles, and each spectrum was normalized using this fit.  The data
were then rebinned in uniform heliocentric wavelength bins.
Figure~\ref{fig:HRS-keyphases} shows the normalized {\Ha} core
emission profiles at several key orbital phases.  As the phase varies,
the profile does not simply shift back and forth in velocity, but
rather changes shape.  To the eye, a decomposition into a broad
component and a narrow component moving in anti-phase is suggested.
This is made more evident in Figure~\ref{fig:trailedspec}, which gives
a trailed-spectrogram representation of all the data.

The measured EW of the {\Ha} emission (above the pseudo-continuum) was
roughly constant at $\sim$3.0\,\AA\ in 2004, but was lower in 2007,
averaging 1.7, 2.1, and 1.8\,\AA\ on 2007 April 19, June 8, and June
22 respectively.  To make Figs.~\ref{fig:HRS-keyphases} and
\ref{fig:trailedspec}, we have ``stretched'' each spectrum to achieve
a uniform emission EW of 3.0\,\AA, keeping the continuum level at
unity.  In the trailed spectrogram (Fig.\ref{fig:trailedspec}), each
spectrum has then been made to fill a phase range of 0.05 cycles,
centered on the phase bin (bin width = 0.01 cycles) nearest to the
orbital phase at mid-observation.  Note that data from 2004 and 2007
have been intermingled, and different orbital phases were observed in
different years. The change in EW from one year to the other may
signal a change in the behavior of the profile {\em vs} orbital phase,
so we caution that quantitative conclusions about these distinct
narrow and broad velocity components, based on these figures, are
tentative.

The phase of the broad component in the {\Ha} emission profile is
given by the RV ephemeris derived in \S\ref{sec:RV}, with the narrow
component moving in anti-phase.  Thus, the Gaussian-derivative method
using FWHM = 11.5\,\AA\ described in \S\ref{sec:RV} likely gives a
``diluted'' RV semiamplitude for the broad component.  Given the
possible change in the profile behavior between 2004 and 2007 noted
above, we did not attempt to decompose each observed profile or
otherwise make formal two-component fits to the high-resolution data.
A rough estimate gives FWHM $\sim$8--10 and $\sim$3\,\AA\ for the
broad and narrow components. Estimates of the RV semiamplitudes are
$K_{\rm broad}\sim 120$ and $K_{\rm narrow}\sim 90$~km~s$^{-1}$, with
errors of perhaps 20~km~s$^{-1}$.

We tentatively identify the broad emission as arising from an
accretion disk around a compact object and the narrow emission as
arising from the mass donating star in a cataclysmic binary.  
If the narrow emission arises on the hemisphere of the mass donor that
faces the disk (see \S\ref{sec:novalike}), one might expect the EW of
the narrow component to vary, such that the emission EW is largest
when the mass donor is at superior conjunction, which is orbital phase
0.5 in our convention.  We searched for this effect using the
low-resolution MDM spectra from the most extensive data set (2005
July), but did not find any firm indication of an EW variation with
orbital phase.  The limit on the variation in the total EW is 15--20\%
full amplitude.  It is not possible to decompose these low-resolution
spectra into separate components.  The high-resolution HRS spectra are
unfortunately not suitable for this test, since all data collected
near orbital phase 0.5 come from the 2007 observations, and all data
colleced near orbital phase 0.0 come from 2004.  While there is a
tantalizing difference in total EW between these two observing
seasons, we cannot know whether it is related to orbital phase; a more
concentrated observing campaign is needed.  We note that there need
not be a large modulation of the EW of the narrow component, if the
orbital inclination of the binary is moderate.

The {\Hb} emission line profile shows behavior similar to {\Ha}, with
the narrow component being perhaps more pronounced (slightly higher
intensity compared with the broad component).  Similar behavior is
also reflected in the \ion{He}{1} $\lambda$5876 emission profile,
although it is weaker.


\subsection{Photometry} \label{sec:Photometry}

\subsubsection{Michigan State University 0.6m Telescope}\label{sec:MSU}

Differential $V$ band photometry of {\LSIV} was obtained using the
Michigan State University 0.6m telescope equipped with an Apogee Ap47p
CCD camera.  Aperture photometry for {\LSIV} was obtained relative to
the star TYC~$5642$$-00482$$-1$, which has $B_T = 12.389$, $V_T =
11.183$.  (This star is significantly redder than the program star,
but there is no blue star of comparable brightness to the target
within the CCD field of view.)  Observations were made on twelve
nights: UT 2004 July 26; 2004 August 1, 7, 9, 16, 22, 31; 2004 September
18; 2005 July 15, 20, 23; and 2006 May 27.  Exposure times were
typically 30 s, varying somewhat depending on observing conditions.
Observing run lengths varied from as short as about an hour to 6.3 hr
on 2006 May 27, this being about the maximum run length possible at
reasonable airmass from the latitude of the observatory. Calibrations
were done using standard procedures and sky flats.  Since observations
were only taken in the $V$ filter, no transformations to the standard
system involving color terms could be attempted.  The typical error
of a single observation is $\pm 0.02$~mag.

\begin{figure}
\epsscale{0.60}
\plotone{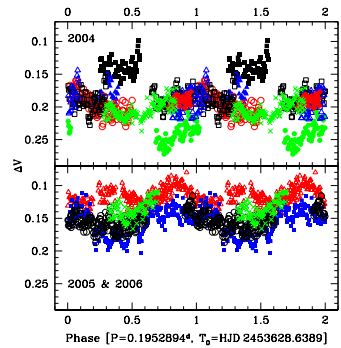}
\caption{Folded differential light curve of {\LSIV}, based on the RV
ephemeris.  The typical error of a single observation is $\pm
0.02$~mag.  Different symbols indicate different observing nights.
Data are plotted twice for phase continuity.  
{\em Top:} observations from 2004 --- 
Jul 26 filled diamond (colored red in electronic edition); 
Aug 1 cross (green); 
Aug 7 filled square (black); 
Aug 9 filled triangle (blue); 
Aug 16 open circle (red); 
Aug 22 open square (black); 
Aug 31 filled circle (green); 
Sep 18 open triangle (blue).
{\em Bottom:} observations from 2005 --- 
Jul 15 cross (green); 
Jul 20 open circle (black); 
Jul 23 filled square (blue) --- and 2006 --- May 27 open triangle (red).
\label{fig:foldedphot}}
\end{figure}

The photometry shows {\LSIV} to be variable on both short (flickering)
and long timescales, with no indication of eclipses.  By itself, the
photometry suggests no clear or consistent period of variation ---
folding the data on numerous trial periods showed many possible alias
periods.  Folding the data on the period derived from the RV
observations (\S\ref{sec:RV}) does show weak modulations or waves,
perhaps multi-periodic, in the brightness of {\LSIV}
(Fig.~\ref{fig:foldedphot}).  The rapid modulation in $\Delta V$ that
is seen in the 2005 Jul 23 data, of amplitude $\sim$0.03 mag and
lasting several cycles over the phase interval $\sim$0.2--0.6, is not
seen in the photometric check stars; this may be simply an instance of
strong flickering. The median light level is about $\sim$0.06 mag brighter
in 2005--2006 than in 2004.  Additional time-series photometry, from a
more southern site to permit longer nightly runs, would be desirable
in characterizing the light curve and searching for persistent
periodic signals.


\subsubsection{Northern Sky Variability Survey}\label{sec:NSVS}

{\LSIV} was detected as a variable star in the Northern Sky
Variability Survey {\citep[NSVS; LS\,IV$-08^\circ 3$ $\equiv$
NSVS~16408817;][]{NSVS}}.  The NSVS record shows 92 observations of
this star.  The typical error of a single measurement is $\sim$0.02
mag.  The data are sparse, collected on 45 nights during the interval
1999 April to 2000 March, with no more than four observations per
night. A light curve folded on our RV period shows no phase-dependent
pattern.  The median unfiltered optical magnitude is 11.855 with a
scatter of 0.052 mag. The total range of observed magnitudes is
11.698$\pm$0.026 to 11.977$\pm$0.068.  The data exhibit variations in
median brightness at the 0.05 mag level with characteristic time
scales of 20--40 days.  Similar to the epoch-to-epoch variations seen
in the MSU observations, this suggests that the system has been
consistent in its photometric behavior over the period of 1999-2006.


\section{Discussion}\label{sec:Discussion}

\subsection{Novalike CV Interpretation}\label{sec:novalike}

We have already suggested (\S\ref{sec:Steward}) that {\LSIV} could
have been misclassified as a hot subdwarf (sdB) star, given only
photometry and spectroscopy confined to wavelengths blueward of {\Hb}.
Indeed, the dwarf nova FO Per (aliases: RL 92 and RWT 92) was classified
by \citet{Chromey79} as an sdB star, when, unbeknownst to him, it must
have been near maximum light in its outburst cycle.  Two Steward
Observatory 2.3m spectra of FO Per, also near maximum light, obtained
in 2004 Dec and 2006 Dec with the identical setup as described in
\S\ref{sec:Steward}, show a spectrum that is almost indistinguishable
from a somewhat reddened sdB star, except for emission cores in
H$\alpha$, H$\beta$, and, very faintly, in the next few higher order
Balmer lines, i.e., at almost exactly the same level as seen in our
2007 May spectrum of {\LSIV}.

The SED of {\LSIV} is considerably redder than those of known single
sdB stars. Dereddening by the equivalent of $E(\bv) \sim 0.6$ or more
would be required to bring the Steward SED at optical wavelengths into
agreement with sdBs that have $T_{\rm eff}$ in the range 24,000 ---
35,000 K.  Likewise, the $E(\bv)$ needed to bring the 2MASS infrared
colors into the range of single sdB stars is 0.6 mag or more.

On the other hand, {\LSIV} shares many properties with novalike
variables of the UX UMa subclass.  These systems show persistent broad
Balmer absorption-line spectra. Ironically, these lines were once
thought to indicate pressure-broadening in the atmosphere of a hot
subdwarf or white dwarf --- e.g., \citet{WH54} --- and only later
were reinterpreted as arising from an optically thick accretion disk,
where they are (at least in part) kinematically broadened.

UX UMa stars also exhibit a wide range in strengths of the emission
line components.  Emission from combinations of {\HI}, {\HeI},
{\HeII}, {\NIII}, and {\CIII} have all been observed at a variety of
strengths in various UX UMa stars, or variable over time in a single
star of the class.  Emission is seen in all of these lines in {\LSIV},
and the features are consistent with those observed in the UX UMa
stars RW Sex and IX Vel
{\citep[][respectively]{BSS92,BeuermannThomas90}}.  For UX UMa stars,
the Balmer decrement in the emission components is steeper than in the
absorption lines.  This can result in a situation where {\Ha} is
purely in emission, while higher members show progressively stronger
(less filled-in) absorption troughs.  This effect is seen in {\LSIV}
(Fig.~\ref{fig:betsyflux2002}).

Narrow Balmer emission has been found in many UX UMa systems,
superimposed on the broader emission from the disk. This narrow
component is attributed to irradiation of the atmosphere of the
secondary star, with reprocessing of the light to Balmer emission.  As
discussed in \S\ref{sec:HRS}, the asymmetry and variations in the
emission line profile shape (Figs.~{\ref{fig:HRS-keyphases}} {\&}
{\ref{fig:trailedspec}}) suggest the presence of such a component,
moving in anti-phase to the broad component, as required for this
interpretation.

The orbital period of $\sim0.195$\,d ($\sim4.7$ hours) determined for
{\LSIV} is comparable to other UX UMa systems.

\subsection{Distance and Reddening}\label{sec:reddening}

{\LSIV} lies at galactic coordinates $(l,b) = (11\fdg0, +20\fdg8)$.
The \citet{Schlegel} reddening maps show that the total line-of-sight
extinction varies on small angular scales in this direction.  We
attempted to infer the $E(\bv)$ color excess directly from the
strength of the DIBs, using the linear relations, $W = a_0 E(\bv)$,
established by {\citet{HerbigDIB}}.  The $\lambda$5780 DIB has an EW
of $0.32 \pm 0.03$\,\AA, suggesting $E(\bv) \approx 0.50$ mag with an
uncertainty of at least 10\%, if we are measuring the narrow component
of this feature only, or $E(\bv) \approx 0.24$ mag if we are measuring
the sum of the narrow and broad DIBs (the latter is more likely). The
$\lambda$4430 DIB has an EW of $0.26 \pm 0.04$\,\AA, corresponding to
$E(\bv) \approx 0.12$ with at least 15\% uncertainty.  Some of the
uncertainty arises, depending on whether ``all-sky'' or ``Sco-Oph''
values of $a_0$ are adopted.  Both DIBs imply reddening less than the
full \citet{Schlegel} column: $E(\bv) = +0.47$.

The heliocentric RV of the interstellar \ion{Na}{1} D lines is
$-14.5~{\rm km~s^{-1}}$ (average of D$_1$ and D$_2$).  Most of this
can be explained by the reflex of the Sun's space motion with respect
to the Local Standard of Rest, so kinematic information about the
distance of the absorbing material between the Sun and {\LSIV} is
unavailable.

If {\LSIV} is assumed to be a UX UMa variable star, how far away is
it, and is this consistent with the measured reddening?  We adopt an
absolute magnitude $M_V \approx +5$ and a representative apparent
magnitude $V = 11.50$ for {\LSIV}. We use the standard relation $A_V =
3.1 E(\bv)$ to compute distance $d$ as a function of assumed $E(\bv)$.
Values between 100 and 200 pc result for $E(\bv)$ in the range 0.0 to
0.5 mag, smaller distances corresponding to larger reddenings.
\citet{Ak07} have put forward a calibration of $M_J$ for CVs, using as
inputs the orbital period and the extinction-corrected 2MASS $J$
magnitude and $J-H$, $H-K_{s}$ colors.  Again varying the assumed
$E(\bv)$, we find $d$ in the range 140--180 pc by this method (minimum
error at fixed reddening is 40\%).  Comparing the 2MASS colors of
{\LSIV} and IX Vel directly, we find a reasonable match if $E(\bv)
\approx 0.10$ mag.  We conclude that, if $E(\bv)$ in the range
0.10--0.20 mag can be accommodated within distances of 100--200 pc,
the colors and magnitudes of {\LSIV} are consistent with those of a UX
UMa variable.  Such reddening values within such distances are
plausible along the line of sight to {\LSIV}, according to the maps
of, e.g., {\citet{PerryJohnston82}}.

\subsection{{\em ROSAT} X-ray Observations}

A weak X-ray source detected in the {\em ROSAT} All-Sky Survey (RASS)
appears to be coincident with {\LSIV}
{\citep[$1$RXS~J$165630.2$$-083442$;][]{Voges}}.  The count rate in
the position sensitive proportional counter (PSPC) was
0.11$\pm$0.02~s$^{-1}$, yielding an estimated 0.1--2.4\,keV flux of
$1.5 \times 10^{-12}~{\rm erg~cm^{-2}~s^{-1}}$.  Taking {\LSIV} to be
the optical counterpart, the X-ray/optical flux ratio is given by
$\log f_X/f_{\rm opt} \approx -2.0$.

This source was also observed serendipitously by {\em ROSAT}
{\citep[$1$WGA~J$1656.4$$-0834$;][]{White}} during a pointed
observation of another target.  In this $\sim$8000\, s exposure, 150
net counts were collected, and the count rate was
0.02$\pm$0.002~s$^{-1}$ (5 times lower than in the RASS),
corresponding to $f_X \approx 3.0\times 10^{-13}~{\rm
erg~cm^{-2}~s^{-1}}$ and $\log f_X/f_{\rm opt} \approx -2.6$.  A
best-fitting blackbody spectrum has $kT \approx 0.2$~keV with an
intervening hydrogen column density of $1.3\times 10^{21}$~cm$^{-2}$,
corresponding to $E(\bv) \approx 0.2$.  The unabsorbed flux is
$F_{\mathrm{0.5-2~keV}} \approx
2.7\times10^{-13}$~erg\,s$^{-1}$\,cm$^{-2}$.  Fits using a bremsstrahlung or
thermal plasma emission model would result in a similar $kT$ value,
given the highly absorbed spectrum and the PSPC's limited energy range
and resolution. The inferred total X-ray flux is quite uncertain.

X-ray emission is not associated with sdB stars.  On the other hand,
CVs of all subclasses show X-ray emission.  {\citet{VBRP97}} discuss
the CVs detected in the RASS.  A typical luminosity in the
0.5--2.5~keV band for a novalike variable is $L_X \sim 10^{31}~{\rm
erg~s^{-1}}$. The required distance for {\LSIV} to have this $L_X$,
given its modeled flux, is $d \sim 500$ pc.  Using a 2 keV
bremsstrahlung spectrum to model the ROSAT counts, {\citet{VBRP97}}
find $\log F_X/F_{\rm UV+opt} < -2.5$ to be typical for novalikes.
Given the uncertainties in the input spectrum and absorption column,
we conclude that {\LSIV} has X-ray properties consistent with
membership in the UX UMa subclass of CVs.

\section{Summary}\label{sec:Summary}

Given the many similarities between our observations and other UX UMa
novalike variables, we propose that {\LSIV} should be classified with
the UX UMa stars.  Evidence supporting this classification includes
the presence of emission reversals in the Balmer absorption series;
other emission lines including \ion{He}{1}, \ion{He}{2}, and
\ion{N}{3}/\ion{C}{3}; the variability of the emission line strengths;
the intrinsically blue continuum; and the mild photometric
variability.  The key evidence in favor of a CV interpretation lies in
the periodically modulated Doppler shifts and profiles of the emission
lines, with a period of 4.7 hr, consistent with a low-mass quasi-main
sequence star that fills its Roche lobe and transfers mass to a
luminous accretion disk around a white dwarf.  The inner hemisphere of
the donor star may be irradiated by the accretion disk.  Reddening and
distance estimates are consistent with the CV interpretation, from the
standpoint of the expected luminosity of the system and the
distribution of interstellar dust in the direction of {\LSIV}.
Finally, weak X-rays are apparently associated with the system, and
the X-ray luminosity and X-ray/optical flux ratio are consistent with
observations of other novalike CVs.

We refrain from deriving the dynamical mass ratio of the binary or
individual masses of the stars in {\LSIV} from any of the various RV
semiamplitudes that we have presented above, mainly because the HRS
spectra on which such estimates would rely were obtained on widely
separated dates, and the EW of the emission lines varied significantly
among the various epochs.  It is thus possible that one or both sites
of the emission varied as well. Work aimed at determining the mass
ratio is best undertaken with intensive time-series spectroscopy that
can provide full orbital phase coverage in one or two nights, to
minimize such variations.  Likewise, we cannot offer a definitive
interpretation of the photometric variability, given our present
limited data.  {\LSIV} thus remains as an attractive and interesting
target for further study.

Despite the apparent line-of-sight extinction to the source, {\LSIV}
is one of the brighter UX UMa stars so far discovered.
{\citet{Downes}} list only three ``UX Uma'' systems (of two dozen
catalogued) that are brighter: IX Vel, V3885 Sgr, and RW Sex.  Only
seven CVs of any ``novalike'' sub-type (89 systems total) are brighter
than {\LSIV} at their maximum light.  Bright, relatively nearby
systems are not only appropriate for follow-up studies, but contribute
disproportionately to studies of the space density and kinematics of
the various subclasses of CVs. Even though hundreds of thousands of
stars that are much fainter have been observed spectroscopically by,
e.g., the Sloan Digital Sky Survey, the sky has not been thoroughly
explored at $V\sim 11$.

\acknowledgments

M.~A.~Stark thanks G.~B.~Berriman, for his help interpreting the 2MASS
catalog, and K.~T.~Lewis, A.~Narayanan, and K.~Herrmann, for assisting
with the observations at KPNO.  We are grateful to K.\ T.\ Lewis for
carrying out the fit to the {\em ROSAT} data, and to L.\ Townsley, who
provided helpful interpretation.
We also thank the HET resident astronomers M.\ Shetrone, S.\ C.\
Odewahn, and H.\ Edelmann and telescope operators F.\ Deglman, M.\
Soukup, M.\ Villarreal, and V.\ Riley.  The Hobby-Eberly Telescope is
a joint project of the University of Texas at Austin, the Pennsylvania
State University, Stanford University,
Ludwig-Maximillians-Universit\"at M\"unchen, and
Georg-August-Universit\"at G\"ottingen. The HET is named in honor of
its principal benefactors, William P. Hobby and Robert E. Eberly.
This research has been supported by grants from NASA (including NASA
GSRP grant NGT5-50399), the Zaccheus Daniel Foundation for
Astronomical Science, and the Pennsylvania Space Grant Consortium.
JRT and CSP gratefully acknowledge funding from the National Science
Foundation through grant AST-0307413, and thank the MDM staff for
expert and conscientious observing support.  Holly Sheets took some of
the MDM spectra.  As always, we are grateful to the Tohono O'odham for
leasing us their mountain so that we may study the great universe
around us.  HAS thanks the Center for Cosmic Evolution and the
National Science Foundation (grant AST-0440061) for their support.
EMG acknowledges support from NSF grant AST-0098699.  This publication
makes use of data products from the Two Micron All Sky Survey, which
is a joint project of the University of Massachusetts and the Infrared
Processing and Analysis Center/California Institute of Technology,
funded by the National Aeronautics and Space Administration and the
National Science Foundation.

{\it Facilities:} \facility{KPNO:2.1m (GoldCam), \facility{Bok (B\&C
spectrograph)}, \facility{Hiltner (modular spectrograph)},
\facility{McGraw-Hill (Mark III spectrograph)}, \facility{HET (HRS)}}


\singlespace

\bibliographystyle{apj}
\bibliography{bibs}

\clearpage



\clearpage

\clearpage

\clearpage

\input{tab1}  
\input{tab2}  
\input{tab3}  
\input{tab4}  
\input{tab5}  

\end{document}

%% file: tab1.tex
\begin{deluxetable}{lll}
\tablewidth{0in}
\tablecaption{Coordinates and photometry of {\LSIV}. \label{table:data}}
\tablehead{
 \colhead{Datum} & \colhead{Value} & \colhead{Reference} }
\startdata
RA (J2000)\dotfill                & 16:56:29.62       & 2MASS\tablenotemark{a} \\ 
Dec (J2000)\dotfill               & $-08$:34:38.7     & 2MASS \\
$\mu_\alpha \cos \delta$ (mas/yr) & $-20.2\pm2.3$     & {\citet{T2}}\tablenotemark{b} \\
$\mu_\delta$ (mas/yr)\dotfill     & $+8.6\pm2.5$      & {\citet{T2}} \\
$V_T$\dotfill                     & $11.564\pm0.114$  & {\citet{T2}} \\
$B_T$\dotfill                     & $11.771\pm0.081$  & {\citet{T2}} \\
$(\bv)_\mathrm{T}$\dotfill        & $+0.207\pm0.140$  & {\citet{T2}} \\
E(\bv)\dotfill                    & $+0.466$          & {\citet{Schlegel}} \\
$V$\dotfill                       & 11.475            & {\citet{RN00}}\tablenotemark{c} \\
(\bv)\dotfill                     & $+0.190$          & {\citet{RN00}}\tablenotemark{c} \\
(\vr)\dotfill                     & $+0.161$          & {\citet{RN00}}\tablenotemark{c} \\
$V_{\mathrm{syn}}$\dotfill        & 11.47             & see \S\ref{sec:RV} \\
$y$\dotfill                       & 11.51             & {\citet{KB92}}\tablenotemark{d} \\
$(b\!-\!y)$\dotfill               & $+0.207$          & {\citet{KB92}}\tablenotemark{d} \\
$m_1$\dotfill                     & $+0.022$          & {\citet{KB92}}\tablenotemark{d} \\
$c_1$\dotfill                     & $+0.072$          & {\citet{KB92}}\tablenotemark{d} \\
$J$\dotfill                       & $10.638\pm0.039$  & 2MASS \\
(\JH)\dotfill                     & $+0.175\pm0.057$  & 2MASS \\
(\JKs)\dotfill                    & $+0.350\pm0.052$  & 2MASS \\
(\HKs)\dotfill                    & $+0.175\pm0.053$  & 2MASS \\
\enddata
\tablenotetext{a}{\citet{Skrutskie06}}
\tablenotetext{b}{{\LSIV} $\equiv$ TYC~5642-627-1.}
\tablenotetext{c}{Based on one measurement.}
\tablenotetext{d}{Based on three measurements.}
\end{deluxetable}

%% file: tab2.tex
\begin{deluxetable}{r@{~}llcccrrc}
\tabletypesize{\small}
\tablewidth{0in}
\tablecaption{Journal of GoldCam Observations \label{table:GoldCamObs}}
\tablehead{
  \multicolumn{2}{c}{UT Date} &
  \multicolumn{3}{c}{Instrument Setup} &
  \colhead{$\lambda\lambda$ range} & Exp.Time & S/N & $\lambda$ for S/N  \\ 
  \cline{3-5} 
  & & \colhead{Grating} & \colhead{Slit ($''$)} & \colhead{Filter} & (s) &
  (\AA) & (pix$^{-1}$) & (\AA)
}
\startdata
2003 Jun &   9 &  \#35\dotfill & 1.5 & OG570 & 6390--8930 &  900                       &  83     & 7560 \\
2003 Jun &  11 &  \#35\dotfill & 1.5 & OG570 & 6390--8930 &  300                       & 120     & 7560 \\
2003 Jun &  14 &  \#26new      & 1.3 & GG385 & 4600--6730 & 1200/1200\tablenotemark{a} & 300/300 & 5665 \\
2003 Sep &  14 &  \#35\dotfill & 1.5 & OG570 & 6390--8930 &  300                       &  73     & 7560 \\
2003 Sep &  15 &  \#35\dotfill & 1.5 & OG570 & 6390--8930 & 300/600\tablenotemark{b}   & 102/145 & 7560 \\
2003 Sep &  17 &  \#26new      & 1.3 & GG385 & 4600--6730 &  450                       & 174     & 5665 \\
2003 Sep &  19 &  \#26new      & 1.3 & GG385 & 4600--6730 &  450                       & 125     & 5665 \\
\enddata
\tablenotetext{a}{Two exposures were taken. The mid-exposure times were separated
by 76 minutes.}
\tablenotetext{b}{Two exposures were taken. The mid-exposure times were separated
by 25 minutes.}
\end{deluxetable}

%% file: tab3.tex

\begin{deluxetable}{lp{6em}rllrrlcr}

\tablewidth{0in}
\tablecaption{Equivalent Widths of Dominant Lines in \LSIV\ from GoldCam Observations. \label{table:EWLSIV}}
\tablehead{
  \colhead{Feature} & \colhead{$\lambda$ (\AA)} &
  \colhead{$\langle\mathrm{EW}\rangle$\tablenotemark{a}} &
  \colhead{$\sigma$\tablenotemark{b}} &
  \colhead{S.D.\tablenotemark{c}} & \colhead{Min\tablenotemark{d}} &
  \colhead{Max\tablenotemark{d}} &
  \colhead{$\Delta$\tablenotemark{e}} & \colhead{N\tablenotemark{f}} &
  \colhead{FWHM\tablenotemark{g}}
}


\startdata
\NIII/\CIII\dotfill                   & $\sim$4650\dotfill    & $-0.52$ & 0.07 & 0.08 & $-0.63$ & $-0.45$ & 0.17 & 4 & 11.90 \\
{\HeII}\dotfill                       & 4685.7\dotfill        & $-0.35$ & 0.06 & 0.04 & $-0.40$ & $-0.31$ & 0.09 & 4 & 8.78  \\
{\Hb} total\dotfill                   & 4861.3\dotfill        & $ 5.21$ & 0.11 & 0.29 & $ 4.82$ & $ 5.52$ & 0.70 & 4 & 26.15 \\
{\Hb} em\dotfill                      & 4861.3\dotfill        & $-0.54$ & 0.05 & 0.09 & $-0.61$ & $-0.43$ & 0.18 & 4 & 7.64  \\
\HeI +\NaI\ D (I.S.)\tablenotemark{h} & 5876\dotfill          & $ 1.56$ & 0.08 & 0.25 & $ 1.32$ & $ 1.80$ & 0.49 & 4 & 7.09 \\
{\HeI} total\tablenotemark{h}\dotfill & 5875.6\dotfill        & $ 0.62$ & 0.06 & 0.17 & $ 0.44$ & $ 0.81$ & 0.37 & 4 & 1.29 \\
{\HeI} em\dotfill                     & 5875.6\dotfill        & $-0.33$ & 0.05 & 0.06 & $-0.38$ & $-0.27$ & 0.11 & 4 & 6.68  \\
{\Ha} total\dotfill                   & 6562.8\dotfill        & $ 0.66$ & 0.22 & 0.76 & $-0.33$ & $ 1.68$ & 2.00 & 8 & 8.50  \\
{\Ha} em\dotfill                      & 6562.8\dotfill        & $-2.60$ & 0.13 & 0.33 & $-3.03$ & $-2.16$ & 0.87 & 8 & 10.38 \\
{\HeI}\dotfill                        & 6678.1\dotfill        & $-0.21$ & 0.09 & 0.04 & $-0.29$ & $-0.16$ & 0.13 & 8 & 6.30  \\
{\OI}\tablenotemark{i}\dotfill        & 7774\dotfill          & $ 0.81$ & 0.13 & 0.13 & $ 0.71$ & $ 0.99$ & 0.28 & 4 & 7.95
\enddata
%
\tablenotetext{a}{Average of all EW measurements (\AA), negative values indicate emission.}
\tablenotetext{b}{Formal error on the average EW (\AA), does not include any systematic errors.}
\tablenotetext{c}{Standard deviation of the measurements of the EW (\AA).}
\tablenotetext{d}{Minimum/maximum EW value measured (\AA).}
\tablenotetext{e}{Difference between the maximum value and minimum value of EW measured (\AA).}
\tablenotetext{f}{Number of individual measurements of the line.}
\tablenotetext{g}{Average FWHM of the feature (\AA).}
\tablenotetext{h}{{\HeI} 5875.6~{\AA} and {\ion{Na}{1}} 5889.9, 5895.9~{\AA}
 are significantly blended; measurements of these lines are very
 uncertain.}

\tablenotetext{i}{Blend of individual components at 7772.0, 7774.2, and 7775.4~\AA.}
\end{deluxetable}

%% file: tab4.tex
\begin{deluxetable}{lrccc}
\tabletypesize{\scriptsize}
\tablewidth{0pt}
\tablecolumns{5}
\tablecaption{Journal of MDM Observations \label{table:mdmjournal}}
\tablehead{
\colhead{Date} &
\colhead{$N$} &
\colhead{HA start} &
\colhead{HA end} &
\colhead{Telescope} \\
\colhead{(UT)} &
\colhead{ } &
\colhead{[hh:mm]} &
\colhead{[hh:mm]} &
\colhead{ }
}
\startdata
2005 Jul 02  &  2  &  $-$1:41 &  $-$1:04 & 1.3m\\
2005 Jul 03  &  1  &  $-$1:26 &  $-$1:26 & 1.3m\\
2005 Jul 07  &  5  &  $-$1:35 &  $-$1:14 & 1.3m\\
2005 Jul 08  &  1  &  $-$1:44 &  $-$1:44 & 1.3m\\
2005 Jul 09  &  1  &  $-$1:40 &  $-$1:40 & 1.3m\\
2005 Jul 10  &  1  &  $-$1:48 &  $-$1:48 & 1.3m\\
2005 Jul 11  &  9  &  $-$1:46 &  +3:11   & 1.3m\\
2005 Jul 12  &  22 &  $-$1:46 &  +3:46   & 1.3m\\
2005 Sep 03  &  5  &  +1:27   &  +3:29   & 1.3m\\
2005 Sep 07  &  3  &  +1:29   &  +1:37   & 1.3m\\
2005 Sep 08  &  2  &  +3:19   &  +3:35   & 1.3m\\
2005 Sep 09  &  9  &  +1:26   &  +2:27   & 1.3m\\
2005 Sep 12  &  2  &  +1:27   &  +1:34   & 1.3m\\
2005 Sep 13  &  3  &  +1:29   &  +1:42   & 1.3m\\
2005 Sep 15  &  4  &  +1:38   &  +1:57   & 1.3m\\
2006 Jan 17  &  12 &  $-$3:58 &  $-$3:11 & 1.3m\\
2006 Jan 18  &  11 &  $-$3:47 &  $-$3:01 & 1.3m\\
2006 Jan 21  &  9  &  $-$3:20 &  $-$2:46 & 1.3m\\
2006 Mar 16  &  13 &  $-$2:43 &  $-$1:49 & 1.3m\\
2006 Mar 17  &  10 &  $-$0:47 &  +0:09   & 1.3m\\
2005 Mar 19  &  1  &  $-$1:21 &  $-$1:21 & 2.4m\\
2005 Jun 29  &  2  &  $-$2:14 &  +1:52   & 2.4m\\
2005 Jun 30  &  1  &  $-$2:26 &  $-$2:26 & 2.4m\\
2005 Jul 01  &  1  &  $-$2:21 &  $-$2:21 & 2.4m\\
2005 Sep 03  &  1  &  $-$2:34 &  $-$2:34 & 2.4m\\
2005 Sep 08  &  1  &  $-$1:55 &  $-$1:55 & 2.4m\\
2006 Mar 17  &  1  &  $-$3:13 &  $-$3:13 & 2.4m\\
2006 Jun 20  &  4  &  +0:01   &  +0:10   & 2.4m\\
2006 Jun 22  &  1  &  $-$1:16 &  $-$1:16 & 2.4m\\
2006 Jun 23  &  1  &  +0:43   &  +0:43   & 2.4m\\
2006 Sep 01  &  1  &  +2:30   &  +2:30   & 2.4m
\enddata
\end{deluxetable}

%% file: tab5.tex
\begin{deluxetable}{llrr}
\tablewidth{0in}
\tablecaption{Journal of HET-HRS observations. \label{table:HRS}}
\tablehead{
  \multicolumn{2}{c}{UT at Start of Exposure} &
  \colhead{Exp} &
  \colhead{Phase\tablenotemark{a}} \\
   & &
  \colhead{(sec)} &
}
\startdata
2004 May 24 & 7:24 & 1144\tablenotemark{b} & 0.161 \\
            & 7:45 &  600 & 0.219 \\
            & 7:55 &  600 & 0.256 \\
2004 May 29 & 7:14 &  600 & 0.713 \\
            & 7:30 &  600 & 0.769 \\
            & 7:40 &  410 & 0.800 \\
2004 May 31 & 7:15 &  600 & 0.956 \\
            & 7:25 &  600 & 0.994 \\
            & 7:36 &  468 & 0.027 \\
2004 Jun 1  & 6:52 &  600 & 0.997 \\
            & 7:03 &  600 & 0.035 \\
            & 7:13 &  600 & 0.072 \\
            & 7:24 &  600 & 0.110 \\
2007 Apr 19 & 9:43 &  600 & 0.470 \\
            & 9:53 &  600 & 0.508 \\
            & 10:04&  600 & 0.545 \\
            & 10:14&  600 & 0.583 \\
2007 Jun 8  & 6:32 &  600 & 0.831 \\
            & 6:42 &  600 & 0.869 \\
            & 6:53 &  600 & 0.906 \\
            & 7:03 &  600 & 0.944 \\
2007 Jun 22 & 5:28 &  600 & 0.293 \\
            & 5:39 &  600 & 0.330 \\
            & 5:50 &  600 & 0.368 \\
            & 6:00 &  600 & 0.406 \\
\enddata
\tablenotetext{a}{Heliocentric orbital phase at the mid-point of the
exposure, determined from the RV ephemeris (\S\ref{sec:RV}).}

\tablenotetext{b}{The first two exposures taken on 2004 May 24 were
inadvertently combined at the telescope into a single 1144~s
equivalent exposure. Start time is for the first observation.}

\end{deluxetable}